\documentclass[12pt,a4paper]{article} 
\usepackage{epsfig}
\usepackage{graphics}
\usepackage{subfigure}
\usepackage{a4wide}
\usepackage{amsmath}

\title{Planar Skyrmions: Vibrational Modes and Dynamics}
 \author{B.M.A.G Piette\footnote{email: B.M.A.G.Piette@durham.ac.uk}\,\,\,
     and R S Ward\footnote{email: richard.ward@durham.ac.uk}
 \bigskip
\\Department of Mathematical Sciences,  \\ University of
Durham, \\Durham DH1 3LE}

\newcommand{\ie}{{\it ie.}}
\newcommand{\half}{{\scriptstyle\frac{1}{2}}}
\newcommand{\quar}{{\scriptstyle\frac{1}{4}}}
\newcommand{\RR}{{\bf R}}
\newcommand{\cL}{{\cal L}}
\newcommand{\pa}{\partial}
\newcommand{\ii}{{\rm i}}
\renewcommand{\a}{\alpha}
\renewcommand{\th}{\theta}
\newcommand{\g}{\gamma}
\renewcommand{\o}{\omega}

\begin{document}

\maketitle \abstract{\noindent
We study Skyrmion dynamics in a (2+1)-dimensional Skyrme model.
The system contains a dimensionless parameter $\alpha$, with
$\alpha=0$ corresponding to the O(3) sigma-model.  If two Skyrmions collide
head-on, then they can either coalesce or scatter --- this
depends on $\alpha$ and on the incident speed $v$, and is affected
by transfer of energy to and from the internal vibrational modes of 
the Skyrmions.  We classify these internal modes and compute
their spectrum, for a range of values of $\alpha$.  In particular,
we find that there is a fractal-like structure of scattering windows,
analogous to those seen for kink-antikink scattering in 1+1 dimensions.
    
}

\newpage

\section{Introduction}

In this paper, we investigate aspects of the dynamics of the
planar Skyrme system, in which the field configuration is a
map from $\RR^2$ to $S^2$. In particular, we examine the discrete
vibrational modes of static $N$-Skyrmion solutions.
Such modes correspond to relatively long-lived vibrational states.
They are of importance for semiclassical quantization (see, for example,
\cite{Wal96,BBT97} for the Skyrme case, \cite{WH00} for the planar-Skyrme
case); but our interest lies in their effect on the classical dynamics
of Skyrmions.

A long-standing prototype is the $\phi^4$ system in $1+1$ dimensions.
Here the soliton is a kink (or antikink), which possesses a single internal
oscillatory mode.  If a kink and an antikink approach each other with low
relative speed $v<v_0$, then they form a long-lived `breather' or bion
state \cite{K75}.  (This is not an exact breather, and eventually decays.)
If, on the other hand, the kink and the antikink approach at high speed 
$v>v_1$,
then they bounce off each other, and each escapes to infinity (the collision is
inelastic, with some radiation being emitted).  For intermediate impact speed
$v_0<v<v_1$, there is a fractal-like structure of `reflection windows', with
trapping and reflection alternating \cite{CSW83,BK88,AOM91}.
This can be understood in terms of a resonant energy exchange between the
translational motion of the kink and its internal oscillation \cite{CSW83}.
Also worth mentioning is that the internal mode can in some sense be
`extrapolated' to a non-infinitesimal dynamical process, namely a
kink-antikink-kink collision \cite{MM97}.

In this paper, we shall study analogous features for the (2+1)-dimensional
Skyrme system.  There are several significant differences between this
case and that of kinks.  The first is that two kinks (as opposed to a kink and
an antikink) always repel each other, and there is no static 2-kink solution;
whereas for Skyrmions (with a suitable choice of potential) static
2-Skyrmion solutions do exist.  Consequently, it makes sense to study
Skyrmion-Skyrmion collisions: for head-on collisions, one might expect that
there will be a critical speed $v_0$ such that
\begin{itemize}
  \item for impact speed $v<v_0$, the Skyrmions coalesce to form a
        2-Skyrmion;
  \item for $v>v_0$, the Skyrmions scatter and each escapes to infinity;
\end{itemize}
and the internal modes should play a role in this process.

As we shall see, however, the picture is not quite as simple as this.
The system contains a dimensionless parameter $\alpha$, and the various
dynamical features depend crucially on $\alpha$. We study numerically how
the spectrum of vibrational modes, and the scattering behaviour, vary
with $\alpha$.  For small $\alpha$, internal modes are absent, and the
picture is indeed as suggested above: coalescence for $v<v_0$, and scattering
for $v>v_0$.  But for larger $\alpha$, a rich spectrum of internal modes
appears, and the scattering behaviour also becomes more complex.
In particular, there is a range of $\alpha$ within which one sees a
fractal-like structure of `scattering windows', separated by regions of
coalescence.





\section{The planar Skyrme system}

The Skyrme system in $\RR^{2+1}$ is defined as follows.
Let $x^{\mu} = (x^0,x^1,x^2) = (t,x,y)$ denote the standard space-time
coordinates; indices are raised and lowered using the standard Minkowski
metric (with signature $+--$).  The two spatial coordinates are denoted $x^j$.
The Skyrme field is a unit vector field $\vec\phi=(\phi_1,\phi_2,\phi_3)$,
with $\vec\phi\cdot\vec\phi=1$.  Its space-time and spatial derivatives
are denoted $\pa_{\mu}\vec\phi$ and $\pa_j\vec\phi$ respectively.
The Lagrangian density is
\begin{equation} \label{Lag}
 \cL = \half(\pa_\mu\vec\phi)\cdot(\pa^\mu\vec\phi)
        - \quar\g\Omega_{\mu\nu}\Omega^{\mu\nu} - \half\a V(\phi_3)\,,
\end{equation}
where $\Omega_{\mu\nu}$ is the triple scalar product
$\Omega_{\mu\nu}=\vec\phi\cdot(\pa_\mu\vec\phi)\times(\pa_\nu\vec\phi)$,
$V$ is a function of $\phi_3$, and $\a$ and $\g$ are constants.

The boundary condition at spatial infinity is $\vec\phi\to(0,0,1)$ as
$r\to\infty$, where $r^2=x^2+y^2$.  A configuration satisfying the boundary
condition has an integer winding number, which we denote $N$, and
which is given by
\begin{equation} \label{N}
   N = \frac{1}{4\pi}\int \Omega_{12}\,d^2x\,.
\end{equation}
The static energy $E$ of a configuration is
\begin{equation} \label{En}
 E = \frac{1}{2}\int\left[(\pa_j\vec\phi)\cdot(\pa_j\vec\phi)
        + \g(\Omega_{12})^2 + \a V(\phi_3)\right]\, d^2x\,.
\end{equation}
The quantity $(\gamma/\alpha)^{1/4}$ has units of length, and so we shall
henceforth fix the length scale by taking $\gamma = \alpha$.  So we have
a system depending on the parameter $\a$, as well as on the
function $V(\phi^3)$.  The static energy of a configuration with winding
number $N$ satisfies the Bogomol'nyi bound \cite{IRPZ92,IW01}
\begin{equation} \label{Bog}
 E\geq4\pi|N|\left(1 + \frac{\a}{2}\int_{-1}^1\sqrt{V(\phi)}\,d\phi\right)\,.
\end{equation}

Clearly for finite energy we need $V(1)=0$.  In the asymptotic region
$r\gg1$, the two components $\phi_1$ and $\phi_2$ (which are the analogues
of the three pion fields in the full Skyrme model) satisfy a
Klein-Gordon equation 
where the `pion mass' $m$ is given by $m^2 = -\a V'(1)/2$.  So the frequency
$\o$ of radiation is bounded below by $m$.  Two choices of $V(\phi)$ for
which the corresponding systems have been investigated in some detail
\cite{PSZ95a,PSZ95b,KPZ98,We99,ESZ00} are $V(\phi)=1-\phi$ and
$V(\phi)=1-\phi^2$; we refer to these as Old Baby Skyrme (OBS) and New
Baby Skyrme (NBS), respectively \cite{We99}.
We shall restrict our attention to these two systems, concentrating
especially on the NBS case; note that previous work on semiclassical
quantization \cite{WH00} dealt with the OBS case.


We remark in passing on the $\a\to\infty$ limit, which amounts to deleting
the $(\pa_{\mu}\vec\phi)^2$ term in the Lagrangian \cite{PTZ92}.  This
system may admit stable solitons.  For example, if $V(\phi)=1-\phi$, then
there is a solution which has compact support (\ie\ $\phi^3\cong1$ outside
a disc of finite radius), and which does not saturate the Bogomol'nyi bound
\cite{GP97}.  With $V(\phi)=(1-\phi)^2$, on the other hand, one has a smooth
static rotationally-symmetric solution with
$\phi^3 = 1-2\exp(-r^2/2N)$, which does saturate the Bogomol'nyi bound.
See \cite{PTZ92} for a more general discussion.

The simplest Skyrmion solutions are rotationally-symmetric, or more accurately
O(2)-symmetric.  Letting $r$ and $\th$ represent the usual polar coordinates
on $\RR^2$, we say that a configuration $\vec\phi(r,\th)$ is O(2)-symmetric if
\begin{itemize}
 \item $\phi_3=\phi_3(r)$ with $\phi_3(\infty)=1=|\phi_3(0)|$; and
 \item $\phi_1+\ii\phi_2=F(r)\exp(\ii N\th)$ with $F(r)$ real-valued.
\end{itemize}
Here $N$ is an integer.  Note that we necessarily have $F(\infty)=0=F(0)$.
An equivalent definition is to say that $\vec\phi$ has the `hedgehog' form
\begin{equation} \label{hedgehog}
  \vec\phi(r,\th)= \bigl(\sin(f)\cos(N\th),\sin(f)\sin(N\th),\cos(f)\bigr)\,,
\end{equation}
where the profile function $f = f(r)$ is smooth and real-valued
with $f(\infty)=0$ and $f(0)=K\pi$ for some positive integer $K$.
There exist symmetric solutions with $K>1$, but they are unstable \cite{KPZ98};
so we shall restrict our attention here to the case $K=1$.  In this case,
$N$ is the same as the winding number (\ref{N}) \cite{PSZ95a}.
For the NBS system, there is considerable numerical evidence that, for each
$N$, there is a smooth minimal-energy $N$-Skyrmion solution, and this solution
is O(2)-symmetric.  For $\alpha=1$, the normalized energy $E_N=E/(4\pi N)$
of this $N$-Skyrmion for $1\leq N\leq4$, obtained by numerical minimization,
is as follows: $E_1=2.15$, $E_2=1.91$, $E_3=1.85$, $E_4=1.83$.  (Note that the
Bogomol'nyi bound (\ref{Bog}) is $E_N\geq1.7854$.)  Since $E_N$ is a decreasing
function of $N$, we expect Skyrmions to coalesce; in particular, a low-speed
collision of two 1-Skyrmions will result in a single 2-Skyrmion.


\section{Vibrational modes of the $N$-Skyrmion}

In this section, we study the spectrum of vibrations about
O(2)-symmetric Skyrmion solutions.  The set of all perturbations about
a symmetric configuration is a vector space which is
acted on by O(2), and which therefore decomposes into disjoint subspaces
corresponding to representations of O(2).  We begin, therefore, with a
brief summary of the irreducible representations of O(2) in this context.

The group O(2) is generated by the rotations $\th\mapsto\th+c$, which
make up SO(2); and the reflections $\th\mapsto-\th$.  It has two
1-dimensional irreducible representations $\langle+\rangle$ and
$\langle-\rangle$; and for each positive integer $p$, a 2-dimensional
irreducible representation $\langle p\rangle$, containing two degenerate
modes $\langle p,+\rangle$ and  $\langle p,-\rangle$.  The corresponding
perturbations $\delta\vec\phi$ of $\vec\phi$ have the following form.
\begin{description}
\item[$\langle+\rangle$:]
        $\delta\vec\phi=\bigl(A\cos(N\th),A\sin(N\th),B\bigr)$,
        with $A\sin f + B\cos f=0$;
\item[$\langle-\rangle$:]
        $\delta\vec\phi=\bigl(-A\sin(N\th),A\cos(N\th),0\bigr)$;
\item[$\langle p,+\rangle$:]
       $\delta\phi_1=B\cos[(N-p)\th)]+C\cos[(N+p)\th)]$,\\
       $\delta\phi_2=B\sin[(N-p)\th)]+C\sin[(N+p)\th)]$,\\
       $\delta\phi_3=A\cos(p\th)$, with $(B+C)\sin f + A\cos f=0$.
\item[$\langle p,-\rangle$:]
       $\delta\phi_1=-B\sin[(N-p)\th)]-C\sin[(N+p)\th)]$,\\
       $\delta\phi_2=B\cos[(N-p)\th)]+C\cos[(N+p)\th)]$,\\
       $\delta\phi_3=-A\sin(p\th)$, with $(B+C)\sin f + A\cos f=0$.
\end{description}
Clearly $\langle+\rangle$ consists of perturbations which maintain the
hedgehog form (\ref{hedgehog}): the profile of the Skyrmion
changes, but it remains O(2)-symmetric.  And $\langle-\rangle$ consists
of perturbations which allow $F(r)$ to become complex-valued but preserve its
modulus (in other words, $F$ acquires an $r$-dependent phase).
So $\langle-\rangle$ includes the zero-mode
$\phi_1+\ii\phi_2\mapsto\exp{(\ii\chi)}(\phi_1+\ii\phi_2)$, where $\chi$
is a constant phase angle.  The other zero-modes are the translations in space;
these form a doublet which belongs to the representation $\langle1\rangle$.
Apart from shape modes in $\langle+\rangle$, some other particularly 
significant
positive modes (cf.\ \cite{Wal96, BBT97} for the (3+1)-dimensional Skyrme
case) are `dipole breather' modes belonging to $\langle1\rangle$, which
correspond to the Skyrmion oscillating from one side to the other; and
the `splitting mode' of the 2-Skyrmion into two single Skyrmions, which
belongs to $\langle2\rangle$.

The spectrum of oscillations around a static solution will consist of
the zero-modes mentioned above (angular frequency $\o=0$); a continuum
of radiation modes ($\o>m$); a finite number of negative modes
($\o$ imaginary) if the solution is unstable; and a finite, possibly zero,
number of discrete positive modes ($0<\o<m$).

We have used two, very different, numerical methods to compute the
vibration spectrum.  The first involves deriving the relevant Sturm-Liouville
equation (in $r$) for each of the classes $\langle+\rangle$,
$\langle-\rangle$, $\langle1\rangle$, $\langle2\rangle, \ldots$;
this is then solved using a Chebyshev spectral method.  This method can be used
for unstable as well as stable solutions.  The second method involves a full
(2+1)-dimensional simulation of the field equations, and only works for
vibrations about stable solutions. The procedure is to construct a static
Skyrmion (by relaxation), put in a perturbation, solve the time evolution
for a long time interval, extract a time-series by sampling the field
at some point in space, and finally Fourier-analyse this data.
See, for example, ref \cite{BBT97}, where this method was used for the
(3+1)-dimensional Skyrme system. The results of the two methods are consistent.

\begin{figure}[htb]
\begin{center}
{
\subfigure{
\includegraphics[scale=0.4]{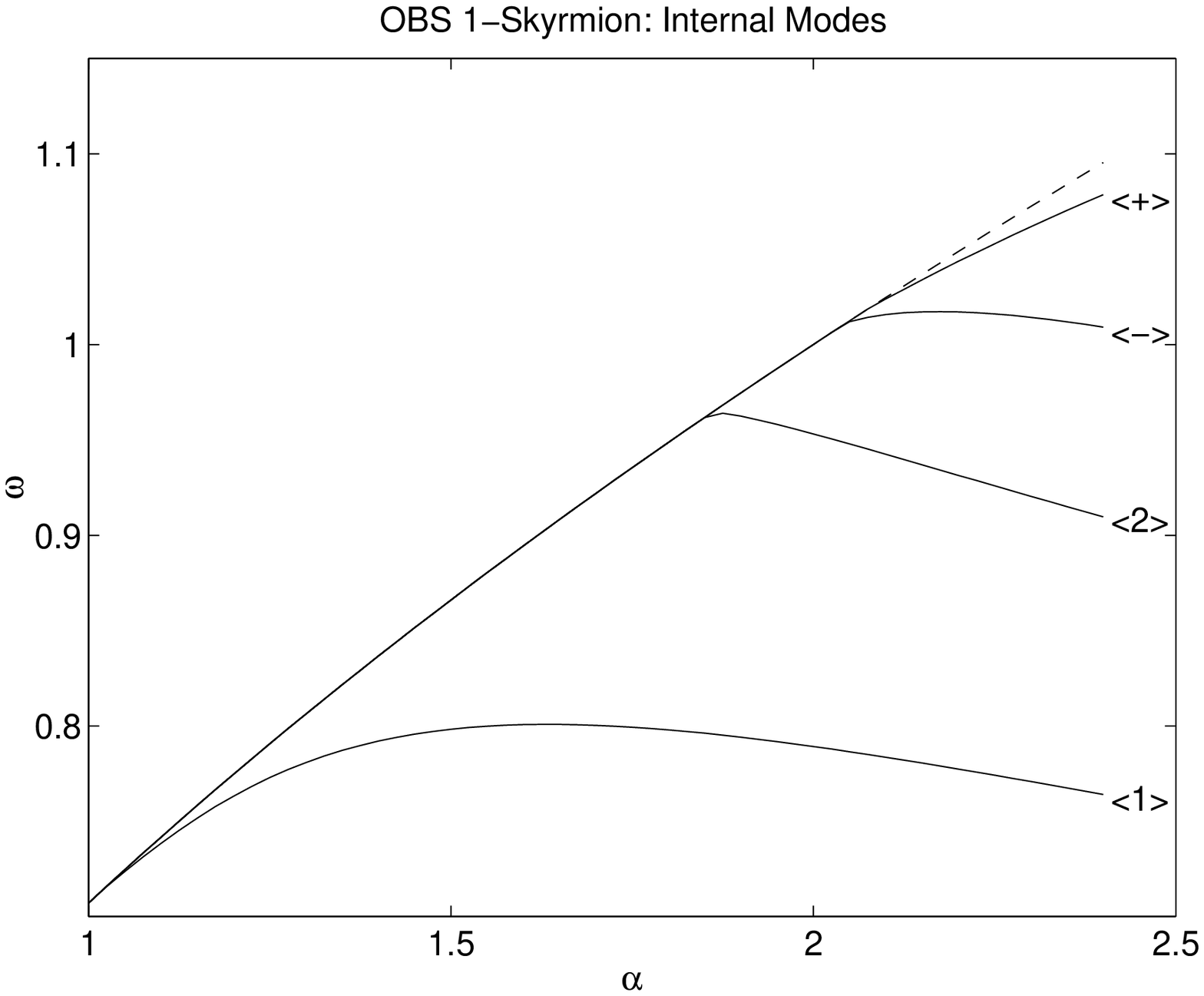}
}
\quad
\subfigure{
\includegraphics[scale=0.4]{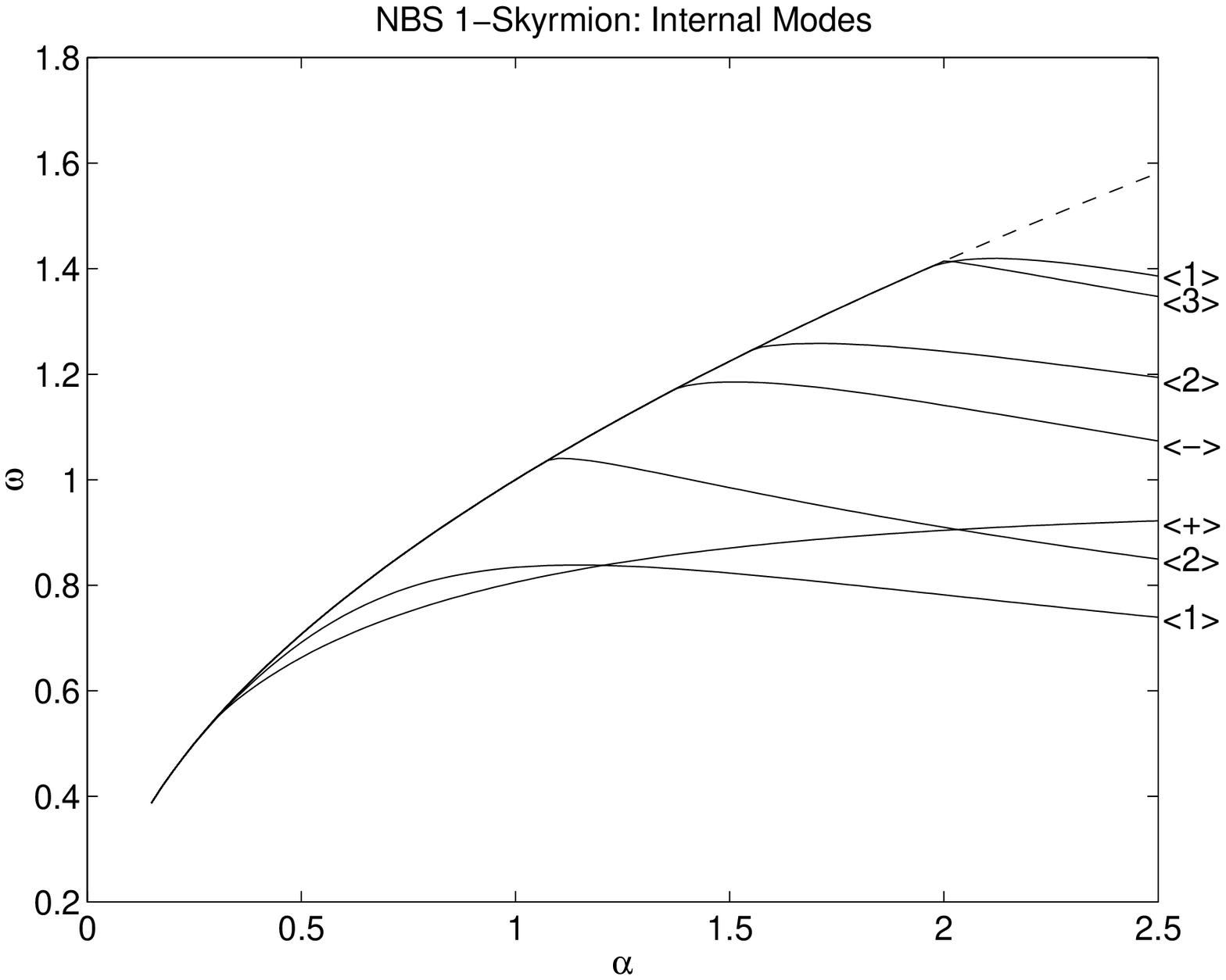}
}
\caption{Frequencies of the internal modes of the 1-Skyrmion.
    \label{fig1}}
}
\end{center}
\end{figure}

Let us now look at the internal modes of the 1-Skyrmion in both the OBS and
NBS systems.  The results are summarized in Figure \ref{fig1}, where
the modes can be seen `peeling off' the bottom of the continuum band
as $\a$ increases.  For the OBS system (where $m^2=\a/2$), there are
no internal modes if $\a<1$.  The first mode to appear belongs to 
$\langle1\rangle$, followed by modes in $\langle2\rangle$, $\langle-\rangle$
and $\langle+\rangle$ in that order.
For the NBS system (where $m^2=\a$), the first mode (again in
$\langle1\rangle$) appears at $\a\approx0.27$, followed by a mode in
$\langle+\rangle$ (at $\a\approx0.30$).  The latter crosses over the
former at $\a\approx0.31$; but at $\a\approx1.2$, the modes cross again, and 
the $\langle1\rangle$ mode has the lower frequency for $\a>1.2$.
The number of modes grows quite rapidly with $\a$.

In the OBS system, the 2-Skyrmion is stable; for all $\alpha>0$, it has a
discrete positive mode in $\langle2\rangle$, corresponding to the breakup of
the 2-Skyrmion into two single Skyrmions (`splitting mode').
As $\a$ increases, more positive modes appear.
The 3-Skyrmion in the OBS system is unstable \cite{PSZ95a}; there is always a
negative mode, which belongs to $\langle2\rangle$.
From now on, we consider the NBS system only.

For the NBS system with $\alpha=1$, the positive discrete modes of the
1-Skyrmion and the 2-Skyrmion are as follows.
\begin{description}
\item[$N=1$:] a shape mode (in $\langle+\rangle$) with $\o=0.806$;
  and a dipole breather doublet in (in $\langle1\rangle$) with
  $\o=0.834$.  See the right-hand diagram in Figure \ref{fig1}.
\item[$N=2$:] a shape mode with $\o=0.519$; a dipole breather doublet
  with $\o=0.613$; a splitting doublet with $\o=0.398$; and two further
  doublets (one in $\langle2\rangle$ with $\o=0.909$ and
  one in $\langle3\rangle$ with $\o=0.912$).  See the right-hand diagram
  in Figure \ref{fig3}.
\end{description}
%



\section{Skyrmion-Skyrmion scattering}

In the NBS system, solitons attract each other, and the lowest-energy
$N=2$ configuration corresponds to two superimposed solitons. This 
means that if two solitons are released from rest at finite separation,
they slowly move towards each other to form a large lump. 
When simulating this numerically, we see that after the scattering, the 
solitons emerge at $90^\circ$, but their mutual attraction slows them down 
and forces them to move back towards each other. They then merge again and 
scatter in their initial direction, but stop again and keep 
oscillating back and forth.  As the interaction is non-elastic, the lumps
emit waves that carry away some energy. As a result, the amplitude of
oscillation slowly decreases, and the 
configuration converges toward the two-soliton static configuration.

If the solitons are sent toward each other with a large-enough speed,
they momentarily merge to form a large lump, and then two lumps emerge at 
$90^\circ$ and move away from each other. If the initial speed is too small,
then these solitons do not escape from their mutual attraction, and they 
eventually form a single lump after several oscillations. 

\begin{figure}[htb]
\begin{center}
{
  \subfigure{
  \includegraphics[scale=0.4]{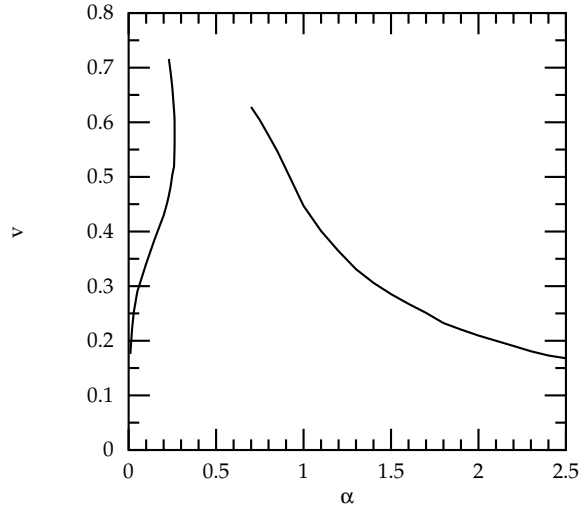}
  }
\caption{Critical velocity as a function of $\alpha$ for the NBS model.
         \label{ScatCV2}}
}
\end{center}
\end{figure}

This behaviour suggests that there is a critical speed below which the
solitons merge, and above which they escape.  In Figure~\ref{ScatCV2}, we
present the dependence of the critical speed $v$ (defined below) on the
coupling constant $\alpha$.
In the limit $\alpha\to0$, the system is the O(3) sigma model,
and a head-on collision always results in $90^\circ$ scattering
\cite{LPZ90,L90}.  In other words, we expect that $v\to0$ as $\alpha\to0$,
and this is consistent with the numerical results.
For small $\alpha$, the Skyrmions scatter very easily; but as $\alpha$
increases towards $0.27$, they have an increasing tendency to coalesce.
In a collision between Skyrmions, any positive discrete (localized) mode 
of the Skyrmion will inevitably be excited (cf.\ \cite{BK88,K01}); this
removes translational kinetic energy from each Skyrmion, and transfers it
to an internal excitation which eventually decays into radiation.
So when the shape mode appears (at $\alpha=0.27$), it very efficiently
removes translational energy from the Skyrmions, and leaves them in a
bound state. There is a partial effect even below $\alpha=0.27$, which can
perhaps be understood in terms of a quasimode (cf.\ \cite{K01}): an `internal'
mode which is embedded within the continuum spectrum, and which becomes
the shape mode when $\alpha$ increases beyond $0.27$.

On the other hand, for $\alpha$ larger than about $0.7$, the critical speed
decreases. Between these two extremes, there is a region where the critical
speed is close to the speed of light ($1$ in our units); but as scattering two 
solitons numerically at large speed is quite difficult, we
were not able to determine the critical speed when $0.27 < \alpha< 0.7$, or
determine for which value of $\alpha$ the $90^\circ$ scattering never takes 
place.

If one looks closely at Figure~\ref{ScatCV2}, one notices that the curve
folds back around the value $\alpha=0.263$. In that region, the solitons merge
when the initial speed is either below the lowest value or above the largest 
one.  The numerical simulations from which Figure~\ref{ScatCV2} was obtained
involve the head-on collision of two Skyrmions in their most attractive
mutual orientation (the `attractive channel'), and with each having
equal initial speed.  The critical speed $v$ plotted in Figure~\ref{ScatCV2}
is defined as the speed below which the solitons
always merge, except for the top part of the folding where it is defined as 
the speed above which the solitons always merge.

When $\alpha$ is slightly smaller than $0.263$, and when
the initial speed is slightly above the lower critical value, we
observed numerically that the scattering behaviour of the solitons can 
have a complex pattern for some values of $\alpha$. 
Instead of simply escaping or merging, the solitons swing back 
and forth as if they were slowly merging, but eventually they 
escape at $90^\circ$ or in the direction from which they came. 

To describe this process, we can define the `scattering number' as the number 
of times the two solitons scatter at $90^\circ$, and define it as zero
when the solitons coalesce. So when the initial speed is very large, the
scattering number is $1$ (except for the narrow range of $\alpha$ around
$0.26$, where high-speed solitons coalesce).
When the scattering number is even, the solitons
scatter in the direction they came from; and when it is odd, they scatter at 
$90^\circ$. For a range of values of $\alpha$, the scattering number is very 
sensitive to the initial velocity, and exhibits a fractal-like structure.

The structure of the data is described in Table~1, where we present the
scattering number for various ranges of initial speed, sampled at regular
intervals.  Each digit in the large strings corresponds to the scattering
number for a given value of $v$. The final string (which is split over two
lines) should be read as a single merged string.  The first digit in each
string corresponds to the smallest value of $v$ (`First $v$' in the table),
while the last digit corresponds to the largest value of $v$ (`Last $v$' in
the table). The increment size in $v$ is given in the third column.


\begin{table}[htb]
\begin{center}
\begin{tabular}{|l|l|l|l|l|}
\hline
$\alpha$ & First $v$ & $v$ increment& Last $v$ & Scattering Numbers \\
\hline
0.24 & 0.4705 & 0.0001 & 0.477 & \\
\multicolumn{5}{|r|}
{002222000000000222005000022000003220000200002002002030200000111111} \\
\hline
$\alpha$ & First $v$ & $v$ increment& Last $v$ & Scattering Numbers \\
0.24 & 0.4721 & 0.00001 & 0.4726 & \\
\multicolumn{5}{|r|}
{222222222222222220000000000000000500000555500000000} \\
\hline
$\alpha$ & First $v$ & $v$ increment& Last $v$ & Scattering Numbers \\
0.24 & 0.472428 & 0.000001 & 0.4725 & \\
\multicolumn{5}{|r|}
{0555555555000000000000000000000000000000000000000700000000000055555555555}\\\hline 
\hline
$\alpha$ & First $v$ & $v$ increment& Last $v$ & Scattering Numbers \\
0.25 & 0.49 & 0.0001 & 0.497 & \\
\multicolumn{5}{|r|}
{22222200000000002222200000000222250000022250000222000020002200200202002} \\
\hline
$\alpha$ & First $v$ & $v$ increment& Last $v$ & Scattering Numbers \\
0.25 & 0.493241 & 0.000001 & 0.49332 & \\
\multicolumn{5}{|r|}
{22222000000000000000005000000000500000005000000000005505555500000040055043333000}\\
\hline 
\hline
$\alpha$ & First $v$ & $v$ increment& Last $v$ & Scattering Numbers \\
0.26 & 0.4905 & 0.0005 & 0.53 & \\
\multicolumn{5}{|r|}
{00002022203002000002222200000002220000022230002200002000220230202022200} \\
\hline
$\alpha$ & First $v$ & $v$ increment& Last $v$ & Scattering Numbers \\
0.26 & 0.5156 & 0.000005 & 0.5159 & \\
\multicolumn{5}{|r|}
{2222222222222222222222222222220000000003000300000003330000000} \\
\hline
$\alpha$ & First $v$ & $v$ increment& Last $v$ & Scattering Numbers \\
0.26 & 0.51578 & 0.000001 & 0.515872 & \\
\multicolumn{5}{|r|}
{0300000000000333400000000000000000333333040000000000000000000000000000}\\
\multicolumn{5}{|r|}
{00005333333333333334040}\\
\hline
\end{tabular}
\caption{Scattering numbers for a range of $\alpha$ and $v$.}
\label{tab1}
\end{center}\end{table}

One clearly sees that for the values of $\alpha$ given in Table~1, there is
no sharp transition between scattering and the merging of solitons, but that
instead there are several windows of initial speed where the solitons scatter
several times before escaping at  $90^\circ$ or in their original directions;
these windows are separated by regions where the solitons do not escape.
As the scattering number increases, the width of the
windows become smaller, but the structure at any scale seems to be similar.
One can also see in Table~1 that often, at the edge of a given 
window, there is a narrower window for a larger scattering number; this is 
not always the case, but we have observed this many times. 

We should also point out that the scattering patterns given in Table~1 form a 
small subset of the range of scattering velocities that we have investigated 
numerically. We have observed a fractal structure in the scattering data for 
the range of speed contained between the lower and upper branch of the 
critical speed on Figure 2 when $\alpha$ was just below $0.26$. 
When $\alpha=0.24$, the windows were not as rich in 
structure as for $\alpha=0.26$, exhibiting large gaps of coalescence between 
the regions of multiple scattering.   


\section{Scattering and Vibration Modes}

Our observations are analogous to those described by Anninos et al
\cite{AOM91} in their study of kink-antikink scattering in the $\phi^4$
model in one dimension, where the scattering also exhibits windows of 
scattering separated by regions where a long-lived bound state is formed.  

\begin{figure}[htb]
\begin{center}
{
\subfigure{
\includegraphics[scale=0.4]{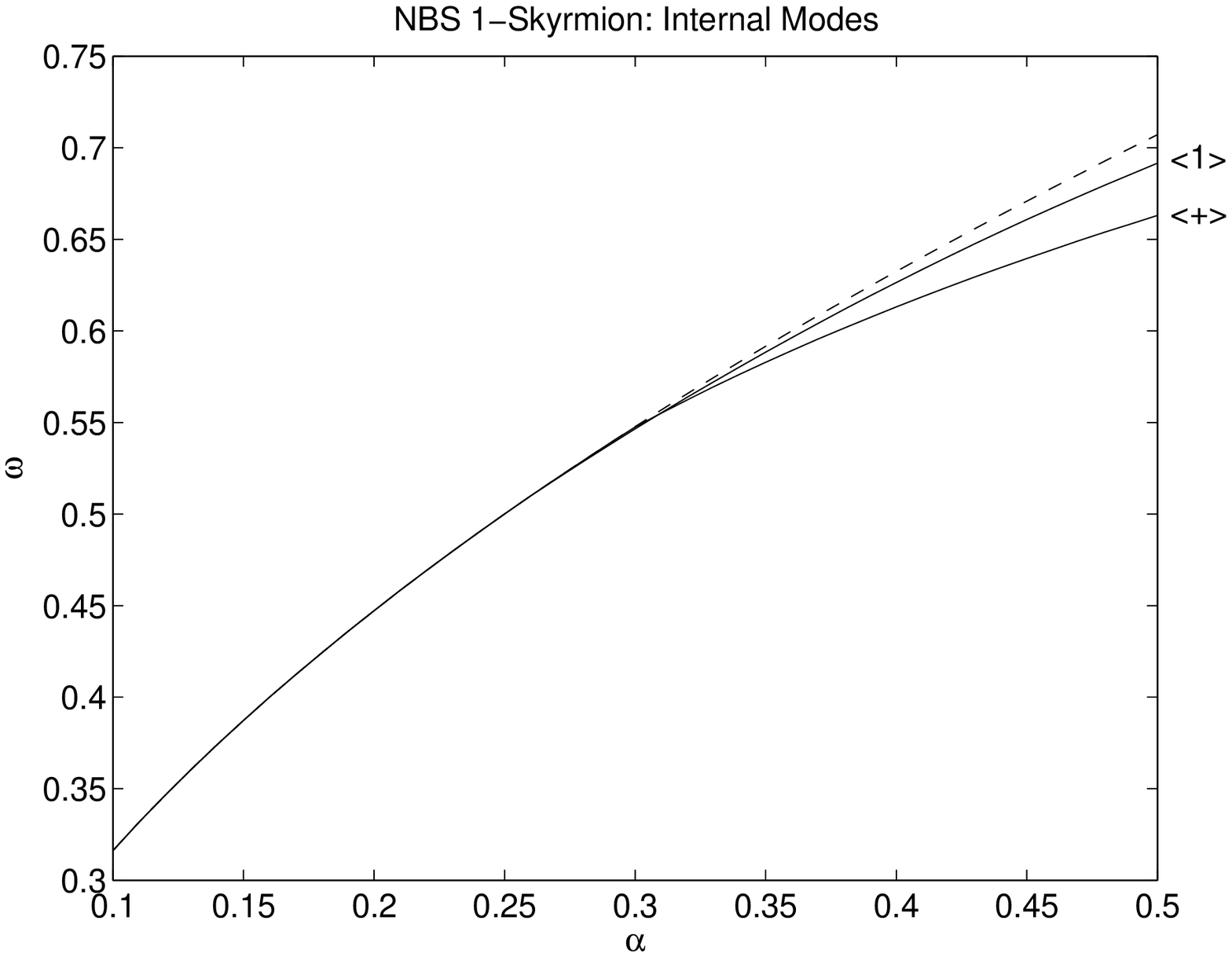}
}
\quad
\subfigure{
\includegraphics[scale=0.4]{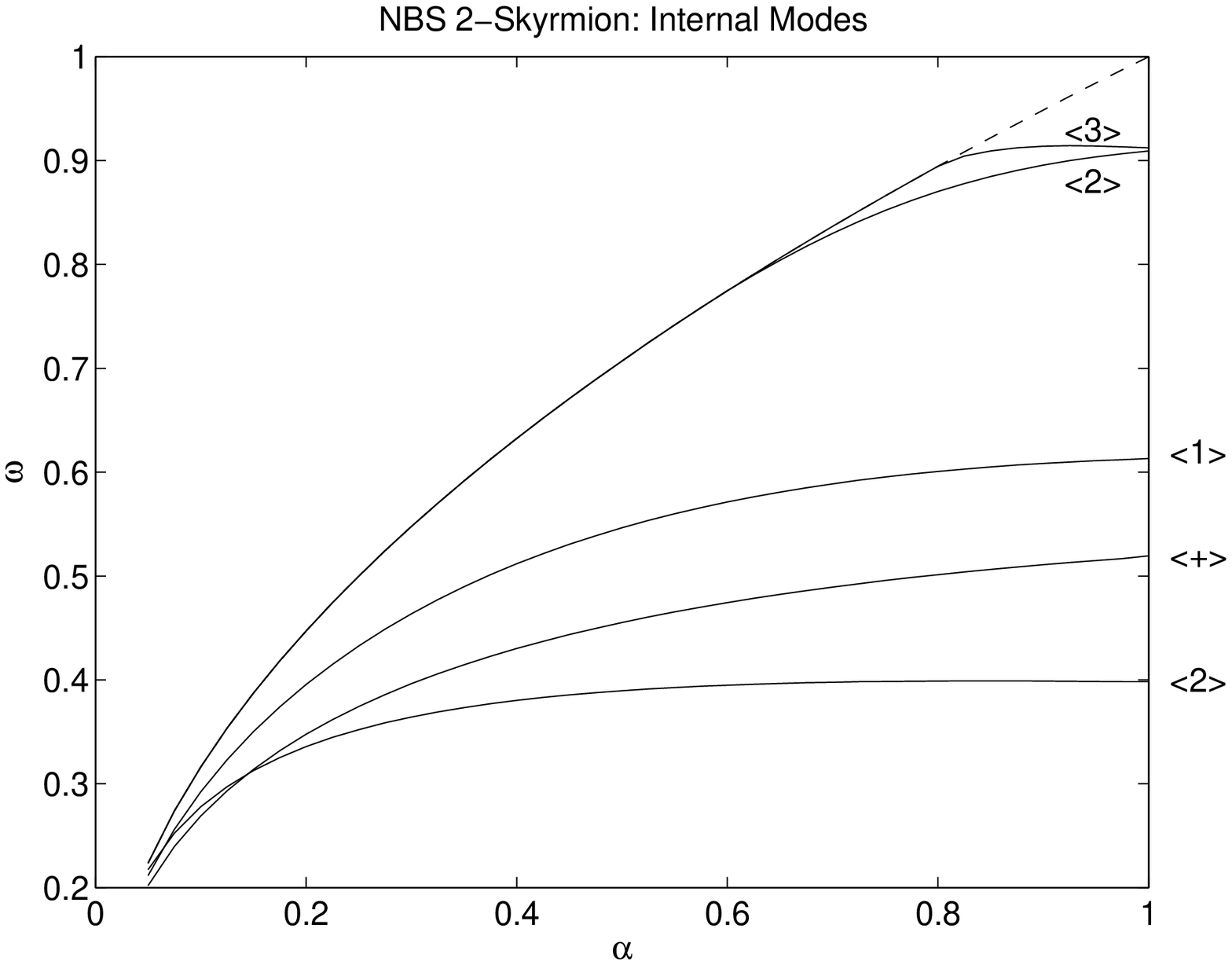}
}
\caption{The frequencies of the internal modes of the $N=1$ and $N=2$ 
Skyrmions.
         \label{fig3}}
}
\end{center}
\end{figure}

It is interesting to note that in Figure~\ref{ScatCV2}, the value
$\alpha=0.263$ where the critical speed curls over is just below the
value $\alpha=0.265$ where the breather mode $\langle1\rangle$ emerges
from the continuum band.  Moreover, the curve for the shape mode
$\langle+\rangle$ emerges at $\alpha = 0.295$, and the curves for the
two modes cross at $\alpha=0.310$.  See Figure~\ref{fig3}.

We also notice from Figure~\ref{fig3} that a splitting mode in 
$\langle2\rangle$
(although not the lowest one) peels off from the continuum band at
$\alpha=0.62$, and this is approximately the value (cf Figure~\ref{ScatCV2})
above which the solitons can scatter, suggesting that this splitting mode is
implicated in the large-$\alpha$ part of Figure~\ref{ScatCV2}.

To understand the relation between the vibration modes and the scattering of 
solitons, we must first realise that during the scattering process the solitons
are excited in various vibration modes. When they are sufficiently far apart, 
each lump can be considered as a single Skyrmion; while when they overlap, 
they form a configuration close to the $N=2$ rotationally-symmetric solution.  
Between these two configurations, they form intermediate states that 
extrapolate between the two extreme configurations, but that are difficult to 
describe. One thing that can be inferred from the expressions for 
$\delta\vec\phi$ is that the mode $\langle p\rangle_1$ for the 1-Skyrmion
becomes the mode $\langle 2p\rangle_2$ for the
2-Skyrmion, as the two Skyrmions merge.  For example, the dipole breather
mode $\langle 1\rangle_1$ for each incident 1-Skyrmion becomes the splitting 
mode $\langle 2\rangle_2$ for the 2-Skyrmion, as the Skyrmions coalesce.
More precisely, the lowest state for $\langle 1\rangle_1$ which is the 
translation mode of the 1-Skyrmion becomes the lowest-state of the 
$\langle 2\rangle_2$ mode. The first excited state of the $\langle 1\rangle_1$
mode, shown in Figures 1 and 3, transforms into the first excited state for the
$\langle 2\rangle_2$ mode, which (as shown in Figure 3) crosses the mass 
threshold around $0.62$. 

Moreover, due to the nonlinear nature of the model, the different vibration 
modes exchange some energy during the scattering. In particular, this means
that some kinetic energy is transfered from the translation mode into the 
various vibration modes, and the reverse is also true when the two
Skyrmions try to escape from their mutual attraction.

During the scattering, the vibration modes that are excited vary with time 
and depend on the overlap between the Skyrmions; but when the initial 
speed is close to the critical value, we observed that the vibration period is 
less than the characteristic time of scattering, implying that the system has
plenty of time to oscillate.

We should also point out that all the scatterings that we have 
simulated numerically involved head-on collisions in the attractive channel
\cite{PSZ95b}, which means in particular that $\vec\phi$ was invariant
under the rotation $\theta\mapsto\theta+\pi$.  This implies that the modes
$\langle p\rangle$ of the 2-Skyrmion, with $p$ odd, were never excited.

We thus see that when $\alpha < 0.265$, there is no internal mode to excite
the 1-Skyrmion, and the oscillation of the Skyrmions induced by their 
scattering is radiated away. When $\alpha > 0.265$, the $\langle 1\rangle_1$ 
is excited during the scattering and it can absorb progressively larger
amounts of energy. As the Skyrmions merge, this mode transforms into the first 
exited $\langle 2\rangle_2$ mode which is well above the mass threshold. The 
energy is thus radiated away and the scattering does not take place. When 
$\alpha > 0.65$, the $\langle 2\rangle_2$ is a bound state, hence the energy 
is not radiated away and the $90^\circ$ scattering becomes possible again.

To try to predict the critical velocity shown in Figure 2, we 
now consider a simple model for the two-soliton scattering process,
namely the dynamics of two equal masses, connected by a spring, on a cone. The 
cone corresponds to the moduli space for the two Skyrmions in their symmetric 
configuration \cite{KPZ93}, while the two masses connected by the spring 
models a single internal mode of oscillation of a Skyrmion. 
Moreover, to model the attraction between the two Skyrmions, the two masses 
are made to evolve in a central potential,
describing the binding energy of two Skyrmions. Each mass is equal to half 
the mass of a Skyrmion, and the masses are positioned so that the line joining
them crosses the origin, ie the tip of the cone.
In this picture, the Skyrmion scatters with its mirror image; and the move 
over the tip of the cone corresponds to a $90^\circ$ scattering once the cone 
is unfolded onto the plane.

The Lagrangian for this system is
\begin{equation}
L = \frac{1}{2} (M_x \dot{X_1}^2 + M_x \dot{X_2}^2) 
        -V_1(X1-X2) - V_2(X_1) - V_2(X_2),
\end{equation}
where
\begin{eqnarray}
V_1(x) &=& \frac{k}{2} (x-X_0)^2\nonumber,\\ 
V_2(x) &=& -\frac{A}{2 \cosh(\lambda x)}.
\end{eqnarray}
The constants
$A$ and $\lambda$ are coefficients, depending on $\alpha$, which
we will choose to model the two-soliton scattering as closely as possible. 
The parameter $X_0$ describes the size of the soliton; we will take 
$X_0=1$. Notice that the depth of $V_2$ is
$A/2$; so given that there are two degrees of freedom, $A$
corresponds to the depth of the two-Skyrmion bound-state potential.
The equations of motion are
\begin{eqnarray}
M_x \ddot{X_1} &=& - k (X_1-X_2) - \frac
   {A \lambda \sinh(\lambda X_1)}{2 \cosh^2(\lambda X_1)},\nonumber \\ 
M_x \ddot{X_2} &=&  k (X_1-X_2) - \frac
   {A \lambda \sinh(\lambda X_2)}{2 \cosh^2(\lambda X_2)}.
\label{EqOneOsc}
\end{eqnarray}
The value of $k$ is chosen so that the frequency of oscillation between $X_1$ 
and $X_2$ is the frequency of the shape mode $\langle+\rangle$, namely
$k/M_x =\omega_{<+>_1}$. The equation
$\omega_{<+>_1}=0.30+0.63(1-\exp(-1.58\ \alpha))$ gives a
good approximation. Note that our simple model does not distinguish between
the different vibration modes; as the frequencies of the $\langle+\rangle$
$\langle1\rangle$ modes are very similar, we have taken the former
for convenience. 

\begin{figure}[htb]
\begin{center}
{ \subfigure{
  \includegraphics[scale=0.4]{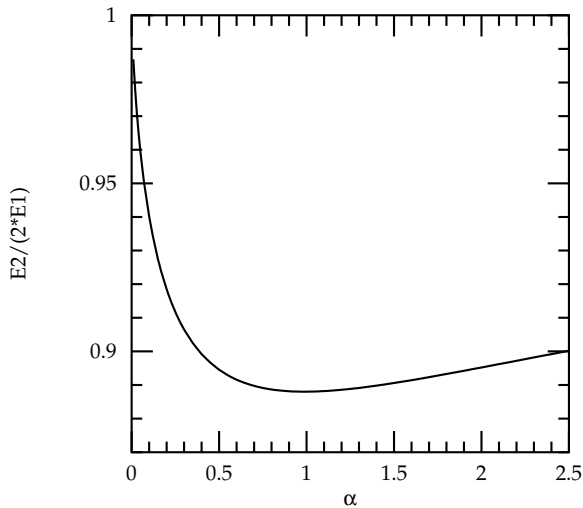}
  }
\caption{$\alpha$-dependence of the binding energy of two Skyrmions.
        \label{EnB1_B2}}
}
\end{center}
\end{figure}

The parameters $A$ and $\lambda$ in the potential $V_2$ are determined
as follows.  For $A$ we take the depth of the potential, namely
$A = 1 -E_2/(2 E_1)$, where $E_2$ is the energy of the two-soliton
bound state and $E_1$ is the energy of a single soliton. 
In Figure~\ref{EnB1_B2} we show the $\alpha$-dependence of the binding energy
$E_2/(2 E_1)$. The curve is well-approximated by the relation
$E_2/(2 E_1)=1+0.12(\exp(-3.88\ \alpha)-1)+0.01\ \alpha$.
%
%
For $\lambda$,
we impose the condition that the frequency of the small-amplitude oscillations 
for $V_2$ is equal to the frequency of the lowest splitting mode for the
two-soliton bound state $\omega_{<2>_2}$. That frequency is well approximated 
by the expression
$\omega_{<2>_2} = 0.2 + 0.2(1-\exp(-3.875\ \alpha))-0.015\alpha$.
We therefore have $\lambda = \omega_{<2>_2}(2 M_{{\rm tot}}/A)^{1/2}$, where 
$M_{{\rm tot}} = 2 M_x$.

To be more realistic, we should have one oscillator for each vibration mode.
One quick way to approximate this is to have $n_0$ decoupled 
oscillators which, to first approximation, have the same frequency.
This will correspond to taking $2n_0$ points of mass $M_x/n_0$ linked in pairs 
by springs with elastic constant $k/n_0$. To preserve the total binding energy 
of the system we must also divide $A$ by $n_0$. This is actually equivalent to
solving equation (\ref{EqOneOsc}), after multiplying $\lambda$ by $n_0^{1/2}$.

To simulate a scattering, we set up the two masses so that they are separated
by their equilibrium distance $X_0$, and so that their centre of mass is 
located at $x=10$. We then send both of them with the same speed towards
the origin. The motion of the masses in the potential well stretches the string
and results in some transfer of translation energy into the
oscillator. The initial speed therefore has to be large enough for the
two masses to go over the tip of the cone and escape towards $x=-\infty$. 
If the speed is too small, the two masses oscillate around the tip of the cone.

\begin{figure}[htb]
\begin{center}
{ \subfigure{
  \includegraphics[scale=0.4]{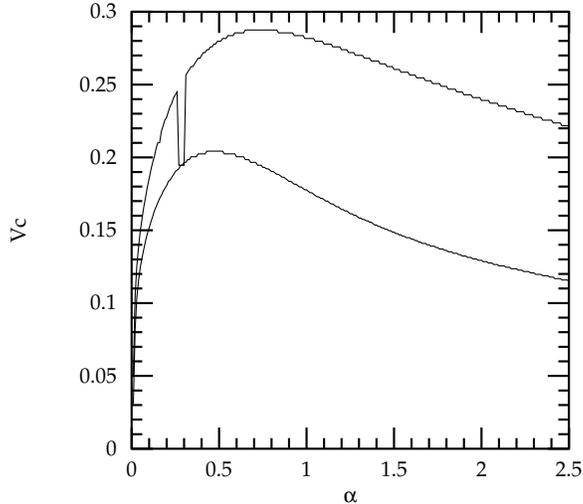}
  }
\caption{Critical speed for the 2-point model; one oscillator (bottom curve) 
  and two oscillators (top curve)
        \label{ModVcNoAbs}}
}
\end{center}
\end{figure}

Using these parameters, we have determined the critical velocity 
for $n_0 = 1$ and $n_0=2$, as shown in Figure~\ref{ModVcNoAbs}.
It produces the correct shape of curve, ie one that looks like the inverse of 
a Morse potential, but the actual critical velocities are too small.
%
%
This is explained by the fact our simple model does not 
take into account the radiation of energy. Otherwise, the shape of the curve 
is more or less explained by the shape of the binding energy of two Skyrmions,
as shown in Figure~\ref{EnB1_B2}.
Where the well is deepest, around $\alpha=1$, more energy is transferred 
into the oscillator, and the critical speed is high.

The drop in the critical velocity for $0.27 \leq \alpha \leq 0.3$
when $n_0 = 2$ is caused by a phase resonance between the oscillation of 
the system in the potential well and the oscillation of the rigid oscillator.
Similar phenomena were also observed when taking a different value for $X_0$.

Figure~1 suggests that our
toy model predicts the critical speed reasonably well for 
large values of $\alpha$, if we take $n_0$ somewhere between $1$ and $2$.
For small values of $\alpha$, the critical speed is too small by 
roughly a factor of 2. The predicted critical speed is also far too small
in the range $0.3<\alpha<0.7$, but the position of the maximum is surprisingly
in the correct region of $\alpha$.

The main success of our simple model is in explaining how the existence
of a critical speed
for $90^\circ$ scattering comes from the fact that some kinetic 
energy is transferred into the vibration modes of the system. It also shows 
that the dependence of the critical speed on $\alpha$ is related to the 
depth of the potential well between the two Skyrmions. To explain the other
features of the curve shown on Figure 2, one must analyse how 
the vibration modes for 1-Skyrmions and 2-Skyrmions transform into one 
another, and consider when these modes are above or below the mass threshold.

When $\alpha < 0.265$, our simple model does not really apply, as the 
1-Skyrmion does not have any genuine vibration mode, although the 2-Skyrmion 
does. When $0.265 < \alpha < 0.62$, the the $\langle1\rangle_1$ mode transforms
into the $\langle2\rangle_2$ which radiates its energy away and makes the
$90^\circ$ scattering difficult or impossible. 

When $\alpha > 0.62$, none of the major modes excited
during the scattering radiates energy away. So some of the energy stored
in these vibration modes can be converted back into the translation mode, 
and the $90^\circ$ scattering can happen at a relatively small speed. 

To predict the critical speed more accurately, one would have to 
take into account all the modes that are excited in the process, as well as 
how they are coupled together, and coupled to the deformation of the system
during the 
scattering. This goes well beyond what we can expect from such a simple model. 
One could consider using a genuine geodesic approximation for the Skyrme 
model, but as we do not have an analytic expression for the general
2-Skyrmions configuration with a separation, this is difficult to do.

The fractal
structure of the scattering data just below $\alpha =0.265$ is more difficult 
to explain. When the vibration modes have a frequency just above the mass 
threshold, the solution for the continuum spectrum exists for all frequencies, 
but for some of them, the eigenfunction has a pronounced maximum at the 
position of the Skyrmion. These solutions can be thought of as oscillation 
modes that decay relatively slowly in the linear limit. These quasi-modes can 
thus absorb some energy and make the system oscillate. The fractal structure 
of the scattering data probably comes from a delicate phase resonance 
between these modes and the scattering oscillations.

\section{Conclusion}
We have shown that the planar Skyrme model possesses some interesting 
vibration and scattering properties. In the limit of small-amplitude
oscillations, the Skymions can vibrate at a finite number of frequencies that 
depend on the coupling constant $\alpha$. 

When two Skyrmion collide head-on, they can coalesce, scatter at $90^\circ$
or, in some very special circumstances, scatter back-to-back. For small values 
of $\alpha$, the Skyrmions can scatter at $90^\circ$ if their collision speed 
is large enough. When $0.265 < \alpha < 0.62$ the $90^\circ$ scattering speed 
seems impossible, within the limit of our numerical integration, while for 
$\alpha > 0.62$ the $90^\circ$ scattering becomes possible again. When
$\alpha$ is just below $0.265$, the scattering data has a fractal structure.

We explained the possibility of $90^\circ$ scattering by studying the
vibration modes of the 1-Skyrmion and 2-Skyrmions solutions.  When the 
deformation mode of the 1-Skyrmion transform into the excited splitting
mode of the 2-Skyrmion, and when this mode is above the mass threshold, a large
amount of energy is radiated and the $90^\circ$ scattering is not possible.
We also described a simple dynamical model which shows that the value of the 
critical speed is related to the binding energy of the two-Skyrmion solution.

\section{Acknowledgement}
This research was supported in part by a research grant from the EPSRC.

\bibliographystyle{plain} 

\end{document}